\documentclass{article}
\usepackage{preprint,times}

\usepackage{amsmath,amsfonts,bm}

\def\eqref#1{equation~\ref{#1}}

\def\1{\bm{1}}

\DeclareMathAlphabet{\mathsfit}{\encodingdefault}{\sfdefault}{m}{sl}
\SetMathAlphabet{\mathsfit}{bold}{\encodingdefault}{\sfdefault}{bx}{n}

\usepackage[colorlinks=true,linkcolor=blue,citecolor=blue,urlcolor=blue]{hyperref}
\usepackage{url}
\usepackage{graphicx}
\usepackage{subcaption}
\usepackage{amsthm}
\usepackage{amsmath}
\usepackage{amssymb}
\usepackage{mathtools}
\usepackage{pifont}
\usepackage{booktabs}
\usepackage{arydshln}
\usepackage{xspace}
\usepackage{algorithm}
\usepackage{algorithmicx}
\usepackage{algpseudocode}
\usepackage{multirow}
\usepackage{wrapfig}
\usepackage{tcolorbox}
\usepackage{mdframed}
\usepackage{framed}
\usepackage{xcolor}
\usepackage{makecell}
\usepackage{array}
\usepackage{bbding}
\usepackage{colortbl}

\definecolor{ForestGreen}{RGB}{64, 136, 39}
\definecolor{redish}{RGB}{212, 57, 57}

\makeatletter
\DeclareRobustCommand\onedot{\futurelet\@let@token\@onedot}
\def\@onedot{\ifx\@let@token.\else.\null\fi\xspace}

\def\eg{\emph{e.g}\onedot} 
\def\ie{\emph{i.e}\onedot} 
 
\def\etc{\emph{etc}\onedot}

\DeclareRobustCommand{\ourapproach}{{\sc KoCo-Bench}\xspace}
\DeclareRobustCommand{\ourapproachbf}{{\sc \textbf{KoCo-Bench}}\xspace}
\makeatother

\theoremstyle{definition}

\theoremstyle{remark}

\usepackage{tcolorbox}
\tcbuselibrary{breakable}
\usepackage{xcolor}

\newtcolorbox[use counter=prompt]{promptbox}[2][]{%
  colback=gray!5,
  colframe=black!50,
  fonttitle=\bfseries,
  title=Prompt: #2,
  breakable,
  #1
}

\iclrfinalcopy
\title{ 
{\sc KoCo-Bench}: Can Large Language Models Leverage Domain Knowledge in Software Development?
}

\author{Xue Jiang$^{1}$, Ge Li$^{1}$, Jiaru Qian$^{1}$, Xianjie Shi$^{1}$, Chenjie Li$^{1}$ \\
\textbf{Hao Zhu$^{1}$, Ziyu Wang$^{1}$, Jielun Zhang$^{1}$, Zheyu Zhao$^{1}$, Lingwei Wu$^{1}$} \\
\textbf{Kechi Zhang$^{1}$, Jia Li$^{2}$, Wenpin Jiao$^{1}$, Zhi Jin$^{1,2}$, Yihong Dong$^{1}$} \\
$^1$ School of Computer Science, Peking University\\$^2$ School of Computer Science, Wuhan University\\
\texttt{\{jiangxue, dongyh\}@stu.pku.edu.cn} \quad \texttt{lige@pku.edu.cn}\\\
}

\begin{document}

\maketitle

\begin{abstract}
LLMs excel at general programming but struggle with domain-specific software development, necessitating domain specialization methods for LLMs to learn and utilize domain knowledge and data. However, \textit{existing domain-specific code benchmarks cannot evaluate the effectiveness of domain specialization methods}, which focus on assessing what knowledge LLMs possess rather than how they acquire and apply new knowledge, lacking explicit knowledge corpora for developing domain specialization methods.
To this end, we present \ourapproach, a novel benchmark designed for evaluating domain specialization methods in real-world software development. \ourapproach contains 6 emerging domains with 11 software frameworks and 25 projects, featuring curated knowledge corpora alongside multi-granularity evaluation tasks including domain code generation (from function-level to project-level with rigorous test suites) and domain knowledge understanding (via multiple-choice Q\&A). Unlike previous benchmarks that only provide test sets for direct evaluation, \ourapproach requires acquiring and applying diverse domain knowledge (APIs, rules, constraints, \etc.) from knowledge corpora to solve evaluation tasks. Our evaluations reveal that \ourapproach poses significant challenges to state-of-the-art LLMs. Even with domain specialization methods (\eg, SFT, RAG, kNN-LM) applied, improvements remain marginal. Best-performing coding agent, Claude Code, achieves only 34.2\%, highlighting the urgent need for more effective domain specialization methods.
We release \ourapproach, evaluation code, and baselines to advance further research at \url{https://github.com/jiangxxxue/KOCO-bench}.  

\end{abstract}

\section{Introduction}

As Large Language Models (LLMs) demonstrate remarkable capabilities in code generation~\citep{claudecode,gpt-5,Gemini3}, the research focus of LLM for Software Engineering (LLM4SE) is shifting from general programming toward domain-specific development~\citep{domcoder,yu2025enhancing}. 
Unlike general programming tasks, domain-specific development demands specialized knowledge, including proprietary rules, processes, API, protocols, dependencies, constraints, etc. The domain knowledge exists in various forms scattered across documentation, source code, or example code. Therefore, how effectively LLMs can acquire, comprehend, and apply domain knowledge is critical to their real-world utility in software engineering.

LLMs exhibit substantial limitations when processing domain knowledge~\citep{domcoder,hung2023walkingtightropeevaluating}, attributable to several factors. First, domain knowledge is inherently specialized, with limited training corpora available, rendering it difficult for models to develop adequate comprehension capabilities through pre-training. Second, the organizational complexity of codebases and data, coupled with fragmented knowledge and implicit semantics, renders models susceptible to incomplete or erroneous understanding during both learning and inference. Moreover, given the rapid evolution of software ecosystems, the inability of LLMs to efficiently acquire and adapt to emerging domain knowledge significantly impairs their effectiveness in practical development contexts.

These limitations have motivated research into domain specialization methods aimed at enabling LLMs to learn and utilize domain knowledge and data. Existing domain specialization methods predominantly rely on general-purpose techniques such as Supervised Fine-Tuning (SFT)~\citep{dong2024abilities,pareja2024unveilingsecretrecipeguide}, Retrieval-Augmented Generation (RAG)~\citep{lewis2020retrieval,gao2023retrieval}, and various extensions thereof~\citep{lora,knn-lm}.
These approaches provide insufficient gains and exhibit notable drawbacks in domain-specific software development. For instance, SFT is constrained by the high cost and scarcity of high-quality domain-labeled data, often leading to overfitting or shallow pattern learning~\citep{ghosh2024closerlooklimitationsinstruction,lin2025debunkmythsftgeneralization}. RAG improves factual recall but struggles with complex reasoning over fragmented and implicit domain knowledge embedded in large codebases~\citep{zhang2025surveygraphrag,agrawal2024mindfulragstudy}. This underscores the urgent need for domain specialization methods that are not only more effective but also tailored to software engineering.

A key obstacle to progress is the lack of code benchmarks for evaluating domain specialization methods. Existing domain-specific code benchmarks are primarily constructed to assess what domain knowledge LLMs already know, rather than how LLMs can better learn and adapt to new domain knowledge. Representative benchmarks, \eg, EvoCodeBench~\citep{li2024evocodebench} and DomainEval~\citep{zhu2024domaineval}, follow a construction pipeline that selects projects from target domains, extracts functions along with corresponding unit tests, and formulates them into code generation tasks with annotated requirements. As a result, these benchmarks consist solely of evaluation test sets and associated code contexts, without an explicitly defined domain knowledge corpus. The absence of knowledge corpus prevents these benchmarks from conducting domain knowledge learning and modeling, confining their utility to performance evaluation and leaving the development of novel domain specialization approaches unsupported.

In this paper, we propose \ourapproach, a novel code benchmark designed to evaluate domain specialization methods for LLMs. Unlike existing benchmarks, \ourapproach innovatively provides knowledge corpora alongside corresponding test sets. 
The benchmark took 28.5 person-months to construct, covering 6 emerging domains: RL, Agent, RAG, Model Optimization, Embodied AI, and Ascend Ecosystem. The knowledge corpus is curated from multiple sources (\ie, framework documentation, framework source code, and usage examples), simulating the introduction of new knowledge sources when LLMs are applied to develop projects based on unfamiliar frameworks. Based on knowledge corpus, we evaluate two tasks: domain code generation and domain knowledge understanding. For domain code generation, we provide multi-granularity requirement specifications (\ie, project descriptions, module divisions, and core function descriptions) to support function-level to project-level generation, with rigorous verification through test suites (\ie, unit tests and integration tests). For domain knowledge understanding, we use easily verifiable multiple-choice Q\&A to evaluate whether LLMs can comprehend the knowledge within the corpus.

We conduct extensive experiments on \ourapproach, encompassing state-of-the-art LLMs (e.g., Claude Sonnet 4.5, Kimi-K2, Gemini 2.5 Pro), representative domain specialization methods (SFT, RAG, kNN-LM), and composite agent systems representing current best practices (Claude Code, SWE-Agent, OpenHands). Beyond standard benchmarking, we perform exploratory studies to investigate how knowledge corpus scale affects learning efficacy, whether continual learning across domains induces catastrophic forgetting, among others. Our experiments yield five key findings: \ding{182} Even SOTA closed-source LLMs struggle with domain code generation and achieve only moderate performance on domain knowledge understanding. \ding{183} Existing domain specialization methods provide only marginal improvements and exhibit inconsistent effectiveness across domains. \ding{184} Agentic approaches currently offer the most effective solution, yet substantial room for improvement remains. \ding{185} Learning-based domain specialization methods show diminishing effectiveness as the knowledge corpus grows larger, and continual learning across domains causes forgetting of previously acquired knowledge.
\ding{186} The most prevalent errors in domain code generation are misusing domain-specific APIs and violating domain data constraints.

\section{Related Work}

\subsection{Code Generation Benchmark}
Existing code generation benchmarks fall into three categories. The first, general coding benchmarks, \eg, HumanEval~\citep{chen2021humaneval}, MBPP~\citep{austin2021mbpp}, LiveCodeBench~\citep{jain2024livecodebench}, primarily test syntax mastery and algorithmic reasoning, often sourced from competitive programming platforms. The second, library-oriented benchmarks, \eg, DS-1000~\citep{lai2022ds1000}, PandasEval and NumpyEval~\citep{zan2022pandaevalnumpyeval}, evaluate the ability of LLMs to understand and apply specific library functions. The third, repository-level benchmarks, \eg, RepoBench~\citep{liu2023repobench}, CrossCodeEval~\citep{ding2023crosscodeeval}, SWE-bench~\citep{jimenez2024swebench}, focus on retrieval and context understanding for code generation and issue resolution. While some recent benchmarks, such as EvoCodeBench~\citep{li2024evocodebench}, DomainCodeBench~\citep{zheng2025domaincodebench}, DomainEval~\citep{zhu2024domaineval}, and MultiCodeBench~\citep{zheng2025domaincodebench}, target domain-specific code generation, they only extract information from repository contexts, thus falling into this third category. 

These benchmarks share a common limitation: they assume models either possess relevant domain knowledge or can extract it directly from given contexts. However, most domain knowledge is neither pre-existing in models nor available in contexts, but rather needs to be learned and understood from domain knowledge corpus. Without an associated knowledge corpus, existing benchmarks cannot support domain knowledge learning and modeling processes, limiting their value to performance evaluation rather than advancing domain specialization methods.

\begin{figure}[t!] 
    \centering 
    \includegraphics[width=0.8\textwidth]{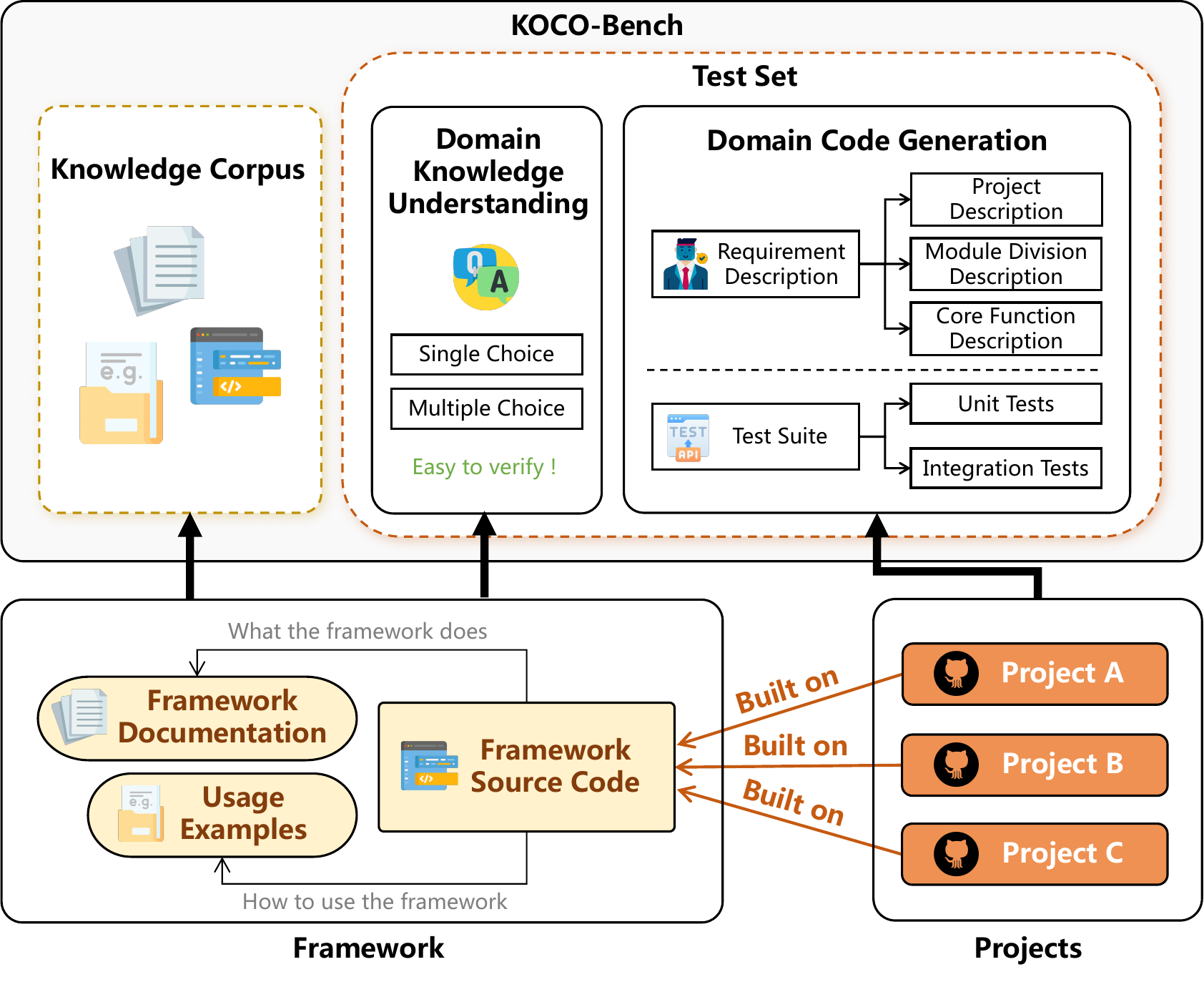} 
    \caption{An overview of \ourapproach.} 
    \label{fig:overview} 
\end{figure}

\subsection{Domain Specialization Method For Software Development}

To adapt LLMs for domain-specific software development, researchers have proposed various domain specialization methods. General approaches include fine-tuning~\citep{dong2024abilities,pareja2024unveilingsecretrecipeguide}, RAG~\citep{gao2023retrieval,lewis2020retrieval}, and kNN-LM~\citep{knn-lm}. Fine-tuning updates model parameters through training on domain data, with parameter-efficient techniques like LoRA~\citep{lora}, QLoRA~\citep{dettmers2023qlora}, and Adapter~\citep{poth2023adapter} developed to reduce computational costs.
RAG retrieves relevant information from external knowledge bases during inference and injects it as context into prompts. kNN-LM retrieves k-nearest neighbors from (hidden state, token) datastores and fuses their output distributions with model predictions for knowledge injection. Researchers have adapted these methods for code domains. kNM-LM \citep{tang2023knm-lm} improves kNN-LM by storing only mispredicted samples and using Bayesian inference for automatic weight calculation. DomCoder \citep{domcoder} combines RAG with Chain-of-Thought reasoning to incorporate domain API knowledge. Other work proposes large-small model collaboration \citep{yu2025enhancing}, where fine-tuned small models provide domain expertise and large models offer general capabilities, with a classifier selecting which generates each token.

However, the evaluation of these methods on software development remains severely inadequate. Existing studies typically use self-constructed naive datasets, training on functions from the same-domain projects or within-project splits, then testing on the remaining functions for code completion.
The incompleteness of this evaluation method is twofold. First, the knowledge overlap is weak among projects within the same domain, or even among functions within a single project, leaving knowledge reference seriously insufficient.
Second, code completion tasks, which predict continuations from given contexts, have multiple valid answers, making evaluation inaccurate.
Without proper consensus benchmarks, despite urgent demand for effective domain specialization methods, this research direction remains underdeveloped.

\section{\ourapproach}

\ourapproach benchmarks LLMs on acquiring and applying external knowledge to software development by simulating the scenario of developing new projects based on domain software frameworks. An overview of \ourapproach is presented in Figure \ref{fig:overview}.

\subsection{Task Formulation}

\ourapproach consists of two core components: Knowledge Corpus and Test Set. The Knowledge Corpus can be utilized through various approaches (e.g., fine-tuning, retrieval augmentation) to enhance the performance of LLMs on the test set. The Knowledge Corpus provides a domain knowledge foundation, including framework documentation, framework source code, and usage examples. The test Set covers the two most common tasks when developers use AI-assisted programming, \ie, domain code generation and domain knowledge understanding: \ding{182} \textbf{The domain code generation task} takes the form of natural language requirement descriptions to code implementation, evaluating the correctness of generated code through test execution. We provide multi-level requirement information (including project descriptions, module division descriptions, and core function descriptions) and multi-level testing (including unit tests and integration tests), supporting code generation evaluation at different granularities from function-level to project-level, to accommodate future trends in code generation technology. \ding{183} \textbf{The domain knowledge understanding task (Q\&A)} can specifically assess models' mastery of particular knowledge points. Existing code repository Q\&A benchmarks \citep{hu2024coderepoqa,liu2021codeqa} typically use GitHub Issues for construction and employ text similarity or LLM-as-judge for evaluation, suffering from the inability to precisely assess domain knowledge and inaccurate evaluation. Our Q\&A task specifically targets domain knowledge assessment and adopts a multiple-choice format that can be precisely evaluated, ensuring the reliability of evaluation.

\begin{figure}[t!] 
    \centering 
    \includegraphics[width=\textwidth]{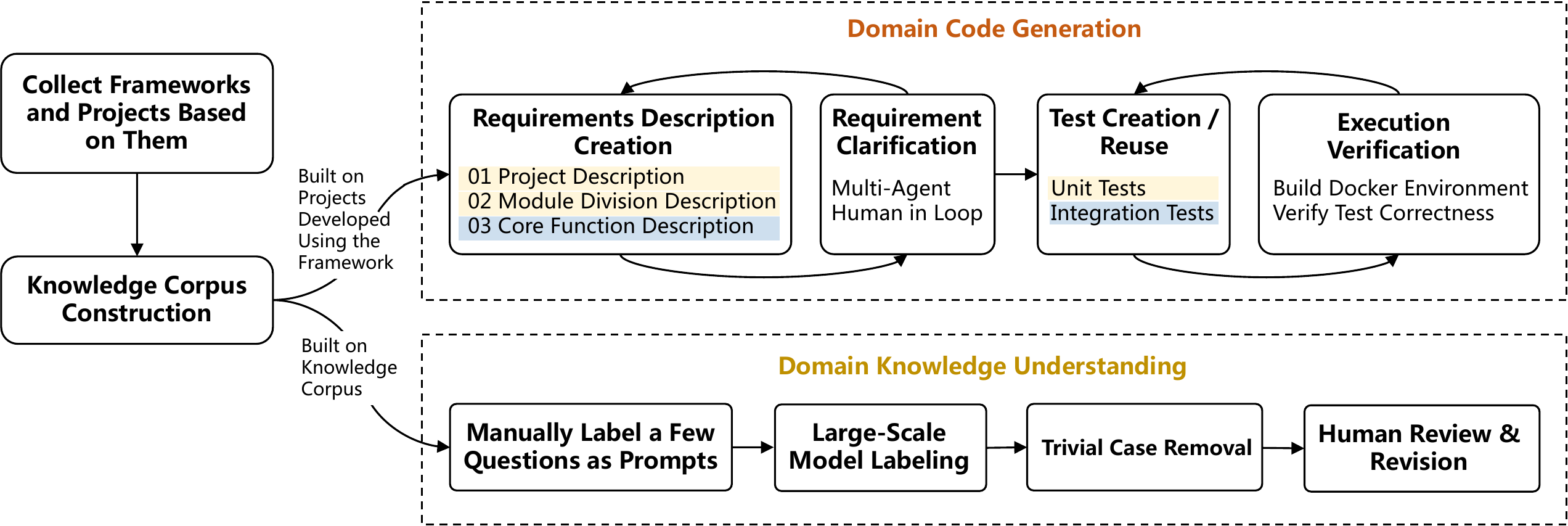} 
    \caption{Construction pipeline of \ourapproach.} 
    \label{fig:construction_pipeline} 
\end{figure}

\subsection{Benchmark Construction}

This section details the construction process of \ourapproach, including the Knowledge Corpus, domain code generation tasks, and domain knowledge understanding tasks. To ensure quality while controlling costs, we follow previous work~\citep{edit-bench} and introduce an LLM agent (\ie, Claude Code) to assist in certain annotation tasks, using it only within work scopes where its performance has been pre-evaluated and confirmed reliable by humans, with all agent-generated content requiring human review and correction. The construction pipeline of \ourapproach is illustrated in Figure \ref{fig:construction_pipeline}.

\subsubsection{Knowledge Corpus}
Constructing a knowledge corpus that naturally aligns with evaluation tasks is typically challenging. We identified software frameworks and their implementations as an ideal scenario for this purpose. Software frameworks not only embody core abstractions, design patterns, and implementation logic but also codify domain knowledge and best practices at the code level. These frameworks usually come with comprehensive documentation and examples, forming essential knowledge components. Developers build various projects based on these frameworks, creating a natural ecosystem of knowledge reuse.
Leveraging this natural knowledge organization, we construct our Knowledge Corpus from three sources: framework documentation, framework source code, and framework usage examples. This corpus then serves as the knowledge foundation for our evaluation tasks. We use projects built with these frameworks to create code generation tasks and design Q\&A tasks to assess knowledge understanding and application.

For corpus construction, we first filter Python-based frameworks from GitHub, selecting those created since March 2024. We further require frameworks to have comprehensive documentation. We then select the top 10 frameworks by GitHub star count, spanning 5 different domains (\ie, RL, Agent, RAG, Model Optimization, and Embodied AI). Additionally, we include two recently open-sourced underlying acceleration and support frameworks for Ascend AI processors (one Python framework and one C++ framework). The frameworks and projects involved are detailed in Table \ref{tab:domain_frameworks} (Appendix).

\subsubsection{Domain Code Generation}

To ensure the completeness of the knowledge corpus, we select projects that are predominantly implemented based on their corresponding frameworks and do not include external dependencies created after the cutoff date. Based on these projects, we construct code generation tasks, involving the creation of requirement descriptions and test suites. This annotation work was completed by 7 annotators over a period of 2 months.

\paragraph{Requirements Description Creation.}

Requirements description is structured at three levels. Project Description provides an overview of project requirements, Module Division Description functions clarify the overall project structure, module responsibilities and interactions, while Core Function Description details the core functional methods of the project. When writing requirements, we ensure they focus on the problem domain rather than code implementation, select core functions related to domain knowledge while excluding utility classes and helper methods (such as config and utils), and avoid redundant or ambiguous non-functional requirements.

To ensure clarity and accuracy of requirements, we employ multi-agent assistance for requirement clarification. First, an agent attempts to write code based on the requirements to identify ambiguous descriptions. This step does not require agents to produce correct code, but rather to expose ambiguous requirements during the implementation process. The identified issues are then fed back to humans and another agent for review and requirement enhancement. This process iterates repeatedly until the requirements are confirmed as clear and complete through human verification.

\begin{table}[t]
\centering
\caption{Statistics of \ourapproach.}
\resizebox{0.8\textwidth}{!}{
\begin{tabular}{lrlrr}
\toprule
 & Total Count & Granular Metrics & \textbf{Mean} & \textbf{Max} \\
\midrule
\multirow{2}{*}{\makecell{Knowledge Corpus\\(Framework)}} 
  & \multirow{2}{*}{11}  & Count per Domain & 1.8 & 4 \\
  &  & Length (Lines) & 77.0k & 400.5k \\
\midrule
\multirow{2}{*}{Projects} 
 & \multirow{2}{*}{25} & Count per Framework & 2.3 & 7 \\
 &  & Length (Lines) & 63.8k & 674.0k \\
\midrule
\multirow{2}{*}{Core Functions} 
 & \multirow{2}{*}{131} & Count per Project & 5.2 & 17 \\
 &   & Length (Lines) & 52.1 & 346 \\
\midrule
\multirow{2}{*}{Tests} 
 & \multirow{2}{*}{978} & Unit Tests per Core Function & 8.6 & 24 \\
 &  & Integration Tests per Project & 2.3 & 6 \\
\midrule
\multirow{2}{*}{QA} 
 & \multirow{2}{*}{107} & Single-Choice Count per Frame & 3.5 & 7 \\
 & & Multi-Choice Count per Frame & 14.3 & 15 \\
\bottomrule
\end{tabular}}
\label{tab:statistics}
\end{table}

\begin{table}[t]
\centering
\caption{Comparison of Different Domain Code Evaluation Benchmarks.}
\resizebox{\textwidth}{!}{
\begin{tabular}{l|c|ccccc|c}
\toprule
\multirow{2}[8]{*}{Benchmark} & \multirow{2}[8]{*}{\begin{tabular}[c]{@{}c@{}}Knowledge\\Corpus\end{tabular}} & \multicolumn{5}{c|}{Domain Code Generation} & \begin{tabular}[c]{@{}c@{}}Domain Code Understanding\end{tabular} \\
\cline{3-8}
 & & \begin{tabular}[c]{@{}c@{}}Project\\Description\end{tabular} & \begin{tabular}[c]{@{}c@{}}Module\\Division\\Description\end{tabular} & \begin{tabular}[c]{@{}c@{}}Core\\Function\\Description\end{tabular} & \begin{tabular}[c]{@{}c@{}}Unit\\Test\end{tabular} & \begin{tabular}[c]{@{}c@{}}Integration\\Test\end{tabular} & Q\&A \\
\midrule
DomainCodeBench & \textcolor{redish}{\XSolidBrush}    & \textcolor{redish}{\XSolidBrush}    & \textcolor{redish}{\XSolidBrush}    & \textcolor{ForestGreen}{\CheckmarkBold} & \textcolor{redish}{\XSolidBrush}    & \textcolor{redish}{\XSolidBrush}    & \textcolor{redish}{\XSolidBrush}    \\
DomainEval & \textcolor{redish}{\XSolidBrush}    & \textcolor{redish}{\XSolidBrush}    & \textcolor{redish}{\XSolidBrush}    & \textcolor{ForestGreen}{\CheckmarkBold} & \textcolor{ForestGreen}{\CheckmarkBold} & \textcolor{redish}{\XSolidBrush}    & \textcolor{redish}{\XSolidBrush}    \\
EvoCodeBench & \textcolor{redish}{\XSolidBrush}    & \textcolor{redish}{\XSolidBrush}    & \textcolor{redish}{\XSolidBrush}    & \textcolor{ForestGreen}{\CheckmarkBold} & \textcolor{ForestGreen}{\CheckmarkBold} & \textcolor{redish}{\XSolidBrush}    & \textcolor{redish}{\XSolidBrush}    \\
MultiCodeBench & \textcolor{redish}{\XSolidBrush}    & \textcolor{redish}{\XSolidBrush}    & \textcolor{redish}{\XSolidBrush}    & \textcolor{ForestGreen}{\CheckmarkBold} & \textcolor{ForestGreen}{\CheckmarkBold} & \textcolor{redish}{\XSolidBrush}    & \textcolor{redish}{\XSolidBrush}    \\
\rowcolor{gray!20} KoCo-bench & \textcolor{ForestGreen}{\CheckmarkBold} & \textcolor{ForestGreen}{\CheckmarkBold} & \textcolor{ForestGreen}{\CheckmarkBold} & \textcolor{ForestGreen}{\CheckmarkBold} & \textcolor{ForestGreen}{\CheckmarkBold} & \textcolor{ForestGreen}{\CheckmarkBold} & \textcolor{ForestGreen}{\CheckmarkBold} \\
\midrule
\end{tabular}}
\label{tab:benchmark_comparison}
\end{table}

\paragraph{Test Suite Creation.}
The test suite comprises unit tests, integration tests, and runtime virtual environments. Unit tests target core functions. If the project already contains unit tests for core functions, we reuse them directly. For functions without tests, we first write test inputs to cover all branches (validating the branch coverage using \texttt{coverage.py}), then run the ground-truth code to obtain expected outputs, and finally have an agent generate the test code. For functions without explicit inputs and outputs, we employ instrumentation to verify the correctness of function call sequences. Integration tests primarily reuse existing pipeline test scripts of the project or are manually written to verify the correctness of multi-module interactions. All tests must pass validation against the ground-truth code and are saved in the corresponding Docker environments.

\subsubsection{Domain Knowledge Understanding}

For frameworks where we could not identify appropriate projects for code generation benchmarks, such as Ascend-related frameworks, we construct domain knowledge understanding Q\&A tasks. This work was completed by 3 annotators over 1.5 months.

The Q\&A task design follows four principles. \ding{182} It only assesses domain-specific knowledge. \ding{183} Answers must be verifiable, so we adopt a multiple-choice format that allows skipping questions and supports multiple selections. \ding{184} Each question maintains atomicity by examining only one point while ensuring questions have unique solutions to avoid ambiguity. \ding{185} Question types are diversified, including complex questions that require synthesizing information from multiple files to answer.

The annotation process begins with humans writing desired, high-quality Q\&A examples as prompts, then using an agent to generate at scale. Next, we filter trivial cases by testing with three LLMs
\footnote{The three models we used are Qwen2.5-Coder-7B-Instruct, Llama-3.1-8B-Instruct, and Deepseek-Coder-7b, whose training data cutoff dates are all before \ourapproach's cutoff date to avoid data contamination.}, eliminating simple questions that all three models answer correctly, as we find these questions typically lack distracting options and are easy to answer. Finally, we conduct a manual review and revision to ensure question quality. Through this rigorous quality control process, we ensure that the Q\&A tasks can effectively evaluate models' understanding of domain knowledge.

\subsection{Features of \ourapproach}

Table~\ref{tab:statistics} summarizes the statistics of \ourapproach, covering the scale and composition of the knowledge corpus, projects, tasks, and evaluation artifacts. Table~\ref{tab:benchmark_comparison} (Appendix) presents a comparison between \ourapproach and existing domain-specific code benchmarks. We highlight four key characteristics:
\ding{182} \textbf{Realistic development scenarios.}
\ourapproach is built from real-world software frameworks and projects. 
The natural alignment between knowledge corpora (frameworks) and evaluation tasks (projects built upon them) ensures that the benchmark reflects software engineering workflow.
\ding{183} \textbf{Diverse domain knowledge.}
\ourapproach covers 6 domains, including 11 knowledge corpora (frameworks) and 25 projects, showcasing substantial diversity. The knowledge corpora average over 76K lines per domain and reach up to 400K lines in the largest case, providing rich domain knowledge for evaluation.
\ding{184} \textbf{Long-context and multi-granularity inputs.}
Projects in test set average more than 63K lines of code, creating long-context scenarios. Together with the knowledge corpus and multi-level requirement specifications, \ourapproach provides diverse inputs, supporting a wide range of methods, including learning-based, retrieval-based, or hybrid methods.
\ding{185} \textbf{Reliable evaluation.}
Unlike benchmarks relying on textual similarity or LLM-based judges, \ourapproach adopts well-defined correctness criteria. Domain Code generation is evaluated via test suites (averaging 8.6 unit tests per function, 2.3 integration tests per project). Domain understanding tasks use single- and multiple-choice questions for unambiguous evaluation.

\section{Experiments}

In this section, we first assess the out-of-the-box performance of current LLMs on \ourapproach. We then examine whether existing domain specialization methods can effectively leverage the knowledge corpus to improve performance. Finally, we evaluate SOTA coding agents that integrate multiple techniques. Beyond these evaluations, we conduct in-depth analyses investigating knowledge forgetting during continual learning and the impact of corpus size on learning effectiveness. Additional experiments on common failure patterns (\S \ref{Structure and Statistics}), project code generation (\S \ref{sec:project_level_generation}), data contamination detection (\S \ref{sec:data_contamination_detection}), and performance-cost trade-offs (\S \ref{sec:performance_cost_trade_offs}), along with case studies (\S \ref{sec:case_study}), and detailed experiment setups (\S \ref{sec:additional_implementation_details}) are provided in Appendix.

\begin{table*}[]
\centering
\caption{Performance of LLMs on \ourapproach across six emerging domains.}
\label{tab:LLMs}
\resizebox{\textwidth}{!}{
\begin{tabular}{llccccccccccccc}
\toprule
\multirow{3}{*}[-10pt]{\makecell{Cutoff\\Date}} & \multirow{3}{*}[-10pt]{Models} & \multicolumn{10}{c}{Domain Code Generation} & \multicolumn{3}{c}{Domain Knowledge Understanding} \\ 
\cmidrule(lr){3-12}\cmidrule(lr){13-15}
 &  & \multicolumn{2}{c}{\makecell{RL \\Frameworks}} & \multicolumn{2}{c}{\makecell{Agent \\ Frameworks}} & \multicolumn{2}{c}{\makecell{RAG \\Frameworks}} & \multicolumn{2}{c}{\makecell{MO\\ Frameworks}} & \multicolumn{2}{c}{\cellcolor{gray!20}Average} & \multicolumn{1}{c}{\makecell{E-AI\\ Frameworks}} & \multicolumn{1}{c}{\makecell{AE\\Frameworks}} & \multicolumn{1}{c}{\cellcolor{gray!20}Average} \\
 \cmidrule(lr){3-4}\cmidrule(lr){5-6}\cmidrule(lr){7-8}\cmidrule(lr){9-10}\cmidrule(lr){11-12}\cmidrule(lr){13-13}\cmidrule(lr){14-14}\cmidrule(lr){15-15}
 &  & Pass@1 & APR & Pass@1 & APR & Pass@1 & APR & Pass@1 & APR & \cellcolor{gray!20}Pass@1 & \cellcolor{gray!20}APR & ACC & ACC & \cellcolor{gray!20}ACC \\
 \midrule
Jan-25 & Gemini-2.5-pro & 7.6 & 23.9 & 9.6 & 21.3 & 0.0 & 0.0 & 16.7 & 21.9 & \cellcolor{gray!20}8.5 & \cellcolor{gray!20}16.8 & 45.7 & 27.0 & \cellcolor{gray!20}36.4 \\
Jul-25 & Claude-Sonnet-4-5 & 7.6 & 28.6 & 5.8 & 21.3 & 0.0 & 0.0 & 11.1 & 34.8 & \cellcolor{gray!20}6.1 & \cellcolor{gray!20}21.2 & 62.7 & 43.8 & \cellcolor{gray!20}53.2 \\
Sep-25* & Kimi-K2-Instruct & 5.6 & 24.3 & 7.7 & 27.6 & 0.0 & 0.0 & 22.2 & 40.6 & \cellcolor{gray!20}\textbf{8.9} & \cellcolor{gray!20}23.1 & 51.7 & 55.4 & \cellcolor{gray!20}\textbf{53.5} \\
Aug-25* & DeepSeek-V3.1 & 5.6 & 20.8 & 7.7 & 27.4 & 0.0 & 0.0 & 11.1 & 34.8 & \cellcolor{gray!20}6.1 & \cellcolor{gray!20}20.7 & 49.5 & 34.7 & \cellcolor{gray!20}42.1 \\
May-24 & GPT-5-mini & 7.8 & 37.4 & 7.7 & 26.3 & 0.0 & 0.0 & 11.1 & 37.6 & \cellcolor{gray!20}6.6 & \cellcolor{gray!20}\textbf{25.3} & 55.8 & 48.6 & \cellcolor{gray!20}52.2 \\
Jun-24 & o4-mini & 5.8 & 22.9 & 5.8 & 18.7 & 0.0 & 0.0 & 16.7 & 29.4 & \cellcolor{gray!20}7.1 & \cellcolor{gray!20}17.8 & 48.5 & 47.7 & 
\cellcolor{gray!20}48.1 \\
\cdashline{1-15}
\multicolumn{15}{l}{\textbf{LLMs trained before KoCoBench $\downarrow$}} \\
Feb-24 & Qwen2.5-Coder-32B-Instruct & 5.8 & 11.4 & 3.9 & 25.0 & 0.0 & 0.0 & 11.1 & 22.7 & \cellcolor{gray!20}5.2 & \cellcolor{gray!20}14.8 & 38.2 & 36.9 & \cellcolor{gray!20}37.5 \\
Feb-24 & Qwen2.5-Coder-7B-Instruct & 3.8 & 11.4 & 3.9 & 17.1 & 0.0 & 0.0 & 22.2 & 33.4 & \cellcolor{gray!20}7.5 & \cellcolor{gray!20}15.5 & 17.6 & 21.6 & \cellcolor{gray!20}19.6 \\
Dec-23 & Llama-3.1-8B-Instruct & 1.8 & 9.4 & 1.9 & 16.6 & 0.0 & 0.0 & 16.7 & 29.4 & \cellcolor{gray!20}5.1 & \cellcolor{gray!20}13.8 & 17.3 & 29.3 & \cellcolor{gray!20}23.3 \\
Feb-23 & Deepseek-Coder-7B & 1.8 & 3.2 & 0.0 & 0.6 & 0.0 & 0.0 & 11.1 & 13.0 & \cellcolor{gray!20}3.2 & \cellcolor{gray!20}4.2 & 5.0 & 2.3 & \cellcolor{gray!20}3.6 \\
\bottomrule
\end{tabular}
}
\begin{flushleft}
\tiny 
1. The six emerging domains include Reinforcement Learning (RL), Agent, Retrieval-Augmented Generation (RAG), Model Optimization (MO), Embodied AI (E-AI), and Ascend Ecosystem (AE). \\
2. The Cutoff Date refers to the training data cutoff of each LLM by default. A Cutoff Date marked with `*' indicates the release date of LLM, as the training data cutoff is not publicly available. \\
3. While baselines score zero in RAG domain, Claude Code's 62.5\% Pass@1 (Table~\ref{tab:detail_agent}) confirms these tasks are solvable via proper domain knowledge access.
\end{flushleft}
\end{table*}

\subsection{Experiment Results}
\paragraph{Performance of Various LLMs on \ourapproachbf.}
We evaluate 10 well-known LLMs on \ourapproachbf, including both closed-source and open-source models across different model families and parameter scales. We explicitly mark each model's training data cutoff date to ensure transparency for potential data contamination. The experimental results are presented in Table \ref{tab:LLMs}.

All LLMs exhibit notably low performance on \ourapproach. Even the best-performing models achieve only single-digit Pass@1 scores (Kimi-K2-Instruct: 8.9\%, Gemini-2.5-pro: 8.5\%), standing in stark contrast to their strong performance on general-purpose benchmarks such as HumanEval, where they exceed 90\% Pass@1. For domain knowledge understanding, even the best LLMs achieve only 53.5\% accuracy. This substantial performance gap underscores the inherent difficulty of domain-specific development and validates the necessity of \ourapproach as a challenging benchmark. Notably, all LLMs score zero on the RAG domain, as it requires domain-specific API calls that none of the models invoked correctly (manually verified). Experiments in Table~\ref{tab:detail_agent} show that Claude Code achieves 62.5\% Pass@1, confirming the task is solvable with proper domain knowledge access. A counterintuitive result emerges in the Model Optimization domain, where larger LLMs sometimes underperform smaller ones. Our analysis reveals that larger LLMs are more prone to invoke external APIs, many of which do not exist or are incompatible with the specified runtime. These results highlight the urgent need for domain specialization methods.

\begin{table*}[]
\centering
\caption{Performance of domain specialization methods on \ourapproach across six emerging domains.}
\label{tab:domain_specialization_methods}
\resizebox{\textwidth}{!}{%
\begin{tabular}{lcccccccccccccc}
\toprule
\multirow{3}{*}[-10pt]{Method} & \multicolumn{10}{c}{\textbf{Domain Code Generation}} & \multicolumn{4}{c}{\textbf{Domain Knowledge Understanding}} \\
\cmidrule(lr){2-11} \cmidrule(lr){12-15}
& \multicolumn{2}{c}{\makecell{RL \\Frameworks}} & \multicolumn{2}{c}{\makecell{Agent \\ Frameworks}} & \multicolumn{2}{c}{\makecell{RAG \\Frameworks}} & \multicolumn{2}{c}{\makecell{MO\\ Frameworks}} & \multicolumn{2}{c}{\cellcolor{gray!20}Average} & \multicolumn{1}{c}{\makecell{E-AI\\ Frameworks}} & \multicolumn{1}{c}{\makecell{AE\\Frameworks}} & \multicolumn{1}{c}{\cellcolor{gray!20}Average} \\
\cmidrule(lr){2-3} \cmidrule(lr){4-5} \cmidrule(lr){6-7} \cmidrule(lr){8-9} \cmidrule(lr){10-11} \cmidrule(lr){12-12} \cmidrule(lr){13-13} \cmidrule(lr){14-14}
& Pass@1 & APR & Pass@1 & APR & Pass@1 & APR & Pass@1 & APR & \cellcolor{gray!20}Pass@1 & \cellcolor{gray!20}APR & ACC & ACC & \cellcolor{gray!20} ACC \\
\midrule
Base Model & 3.8 & 11.4 & 3.9 & 17.1 & 0.0 & 0.0 & 22.2 & 33.4 & \cellcolor{gray!20}7.5 & \cellcolor{gray!20}15.5 & 17.6 & 21.6 & \cellcolor{gray!20}19.6 \\
SFT & 7.3 & 12.9 & 1.9 & 16.1 & 0.0 & 0.0 & 16.7 & 24.4 & \cellcolor{gray!20}6.5 & \cellcolor{gray!20}13.4 & 25.6 & 33.8 & \cellcolor{gray!20}29.7 \\
LoRA & 5.5 & 16.1 & 3.9 & 14.0 & 0.0 & 0.0 & 16.7 & 21.9 & \cellcolor{gray!20}6.5 & \cellcolor{gray!20}13.0 & 17.9 & 35.5 & \cellcolor{gray!20}26.7 \\
RAG & 7.2 & 23.7 & 5.8 & 12.8 & 0.0 & 0.0 & 16.7 & 38.9 & \cellcolor{gray!20}7.4 & \cellcolor{gray!20}18.9 & 25.1 & 35.5 & \cellcolor{gray!20}30.3 \\
kNN-LM & 5.4 & 12.3 & 1.9 & 16.1 & 0.0 & 0.0 & 16.7 & 25.0 & \cellcolor{gray!20}6.0 & \cellcolor{gray!20}13.3 & 21.5 & 29.3 & \cellcolor{gray!20}25.4 \\
\bottomrule
\end{tabular}%
}
\begin{flushleft}
\tiny 
1. The six emerging domains include Reinforcement Learning (RL), Agent, Retrieval-Augmented Generation (RAG), Model Optimization (MO), Embodied AI (E-AI), and Ascend Ecosystem (AE). \\
2. While baselines score zero in RAG domain, Claude Code's 62.5\% Pass@1 (Table~\ref{tab:detail_agent}) confirms these tasks are solvable via proper domain knowledge access.
\end{flushleft}
\end{table*}

\paragraph{Evaluation of Domain Specialization Methods.}

We evaluate various domain specialization methods on \ourapproach. We adopt Qwen2.5-Coder-7B-Instruct as the base model, whose training data cutoff predates our dataset's cutoff, thereby avoiding potential data leakage. Specifically, we assess four representative general methods: SFT, LoRA, RAG, and kNN-LM. These methods differ in how they incorporate domain knowledge: SFT and LoRA are training-based approaches that update model parameters, RAG is an inference-time retrieval method, and kNN-LM combines both training (datastore construction) and inference (retrieval).
The results in Table~\ref{tab:domain_specialization_methods} reveal that existing domain specialization methods struggle with domain code generation. Each method exhibits varying effectiveness across domains, achieving gains in some while incurring losses in others. For domain knowledge understanding, all methods show improvements, yet the best accuracy reaches only 30.3\%. Among all methods, RAG performs best, followed by SFT.
We further evaluate two code-specific methods (KNM-LM and DSCC), originally designed for code completion tasks. We carefully adapt them to code generation, with results in Appendix Table~\ref{tab:code_domain_specialization_methods}. Both methods underperform the base model: Pass@1 shows nearly no improvement, and APR gains are marginal, with performance drops in certain domains. This may be attributed to the difficulty of adapting completion-oriented methods to generation tasks with rigorous test validation.
These findings underscore the urgent need for effective, code-oriented domain specialization methods.

\paragraph{Performance of LLM Agents on \ourapproachbf.}

We also evaluate agent methods that combine multiple techniques (\eg, search, tool use, and memory) to further advance the effectiveness of domain knowledge utilization. We assess two representative open-source coding agents (SWE-Agent \citep{yang2024sweagent} and OpenHands \citep{wang2025openhands}) and one popular closed-source coding agent tool (Claude Code \citep{claudecode}) on the domain code generation task to investigate the performance ceiling achievable by SOTA techniques on \ourapproach. The open-source agents are based on Qwen2.5-Coder-32B-Instruct, while Claude Code is powered by Claude-Sonnet-4-5. All agent methods operate with full access to the project folder and knowledge corpus, performing iterative exploration, code generation, and execution-validation cycles. The results are presented in Table \ref{tab:detail_agent}.

\begin{table*}[t]
\centering
\caption{Performance of open-source and closed-source agents on \ourapproach across six emerging domains.}
\label{tab:detail_agent}
\resizebox{\textwidth}{!}{%
\begin{tabular}{lccccccccccc}
\toprule
\multirow{2}{*}{Method} & \multicolumn{2}{c}{\makecell{RL \\Frameworks}} & \multicolumn{2}{c}{\makecell{Agent \\ Frameworks}} & \multicolumn{2}{c}{\makecell{RAG \\Frameworks}} & \multicolumn{2}{c}{\makecell{MO\\ Frameworks}} & \multicolumn{2}{c}{\cellcolor{gray!20}Average} & \multirow{2}{*}[-3pt]{Token Cost} \\
\cmidrule(lr){2-3} \cmidrule(lr){4-5} \cmidrule(lr){6-7} \cmidrule(lr){8-9} \cmidrule(lr){10-11}
& Pass@1 & APR & Pass@1 & APR & Pass@1 & APR & Pass@1 & APR & \cellcolor{gray!20}Pass@1 & \cellcolor{gray!20}APR & \\
\midrule
\multicolumn{11}{l}{\textbf{Base Model}} \\
Qwen2.5-Coder-32B-Instruct & 5.8 & 11.4 & 3.9 & 25.0 & 0.0 & 0.0 & 11.1 & 22.7 & \cellcolor{gray!20}5.2 & \cellcolor{gray!20}14.8 & 1,247 \\
\cdashline{1-12}
\multicolumn{11}{l}{\textbf{Open-source Agent}} \\
SWE-agent & 3.6 & 6.6 & 1.9 & 12.6 & 0.0 & 0.0 & 12.5 & 22.5 & \cellcolor{gray!20}4.5 & \cellcolor{gray!20}10.4 & 26,583  \\
OpenHands & 1.8 & 4.8 & 0.0 & 0.0 & 0.0 & 2.5 & 12.5 & 14.4 & \cellcolor{gray!20}3.6 & \cellcolor{gray!20}5.4 & 26,005 \\
\midrule
\multicolumn{11}{l}{\textbf{Base Model}} \\ 
Claude-Sonnet-4-5 & 7.6 & 28.6 & 5.8 & 21.3 & 0.0 & 0.0 & 11.1 & 34.8 & \cellcolor{gray!20}6.1 & \cellcolor{gray!20}21.2 & 1,760 \\
\cdashline{1-12}
\multicolumn{11}{l}{\textbf{Closed-source Agent}} \\
Claude Code & 13.6 & 26.4 & 38.5 & 56.3 & 62.5 & 68.8 & 22.2 & 45.6 & \cellcolor{gray!20}34.2 & \cellcolor{gray!20}49.3 & 619,923 \\
\bottomrule
\end{tabular}%
}
\begin{flushleft}
\tiny 
1. The six emerging domains include Reinforcement Learning (RL), Agent, Retrieval-Augmented Generation (RAG), Model Optimization (MO), Embodied AI (E-AI), and Ascend Ecosystem (AE). \\
2. Token Cost denotes the average number of tokens generated per sample, including both input and output tokens, for each baseline.
\end{flushleft}
\end{table*}

Claude Code significantly outperforms SWE-Agent and OpenHands, achieving a Pass@1 of 34.2\%. Notably, Claude Code attains a Pass@1 of 62.5\% in the RAG domain, effectively solving code generation tasks that other domain specialization methods fail to address. We analyze three primary factors contributing to the limited effectiveness of SWE-Agent and OpenHands: 1) the base model (Qwen2.5-Coder-32B-Instruct) has not been specifically trained for agentic coding scenarios, resulting in weaker agentic capabilities; 2) both agents are prone to instruction drift, gradually losing track of the original repo-level task requirements and function signatures during extended interactions; and 3) these agents frequently fall into repetitive loops of tool invocations when attempting to gather contextual information, wasting computational resources without making progress. Moreover, agentic methods suffer from high inference costs, consuming more tokens than other approaches. To address this confound, we conduct control experiments with Claude-Sonnet-4-5 as the base model for all agents (\S \ref{sec:control_agent}). The results show that switching to the stronger base model yields notable improvements with both SWE-Agent and OpenHands, yet their performance still falls substantially short of Claude Code, revealing a gap between current open-source and closed-source agents.

\subsection{Further Analysis} \label{sec:analysis}

\paragraph{Investigating Knowledge Forgetting and Conflict.} 
In real-world deployment, models often continue learning across multiple frameworks or domains. A critical question is whether such sequential knowledge acquisition leads to forgetting or knowledge conflicts that degrade performance on previously learned content. 
We investigate this problem on \ourapproach using SFT, and the results are presented in Table~\ref{tab:frame_domain_performance}. When sequentially training on two frameworks within the same domain, the model maintains comparable performance to the single-framework baseline, with AvgPassRatio even showing a slight improvement. In contrast, cross-domain sequential training leads to notable performance degradation, with Pass@1 dropping from 7.1\% to 3.6\%.
These results indicate that knowledge forgetting or conflict is largely determined by the semantic relationship between training corpora. When frameworks are closely related, joint training can yield synergistic benefits through shared representations. However, when frameworks originate from distant domains, sequential learning is more likely to cause knowledge overwriting, resulting in degraded performance on previously learned content. This finding suggests that practitioners should exercise caution when extending models across domains, and that incorporating data replay strategies during training may be necessary to preserve prior knowledge.

\begin{table}[tbp]
\centering
\caption{Impact of continual learning on knowledge.}
\label{tab:frame_domain_performance}
\resizebox{0.65\textwidth}{!}{
\begin{tabular}{l c c c}
\toprule
 & \makecell{Single\\Framework} & \makecell{Multi-Framework\\Same Domain}  & \makecell{Multi-Framework\\Cross-Domain} \\
\midrule
Pass@1 & 7.1 & 7.1 & 3.6 \\
AvgPassRate & 12.8 & 13.5 & 11.1 \\
\bottomrule
\end{tabular}
}
\end{table}

\begin{figure}[h!] 
    \centering 
    \includegraphics[width=0.55\textwidth]{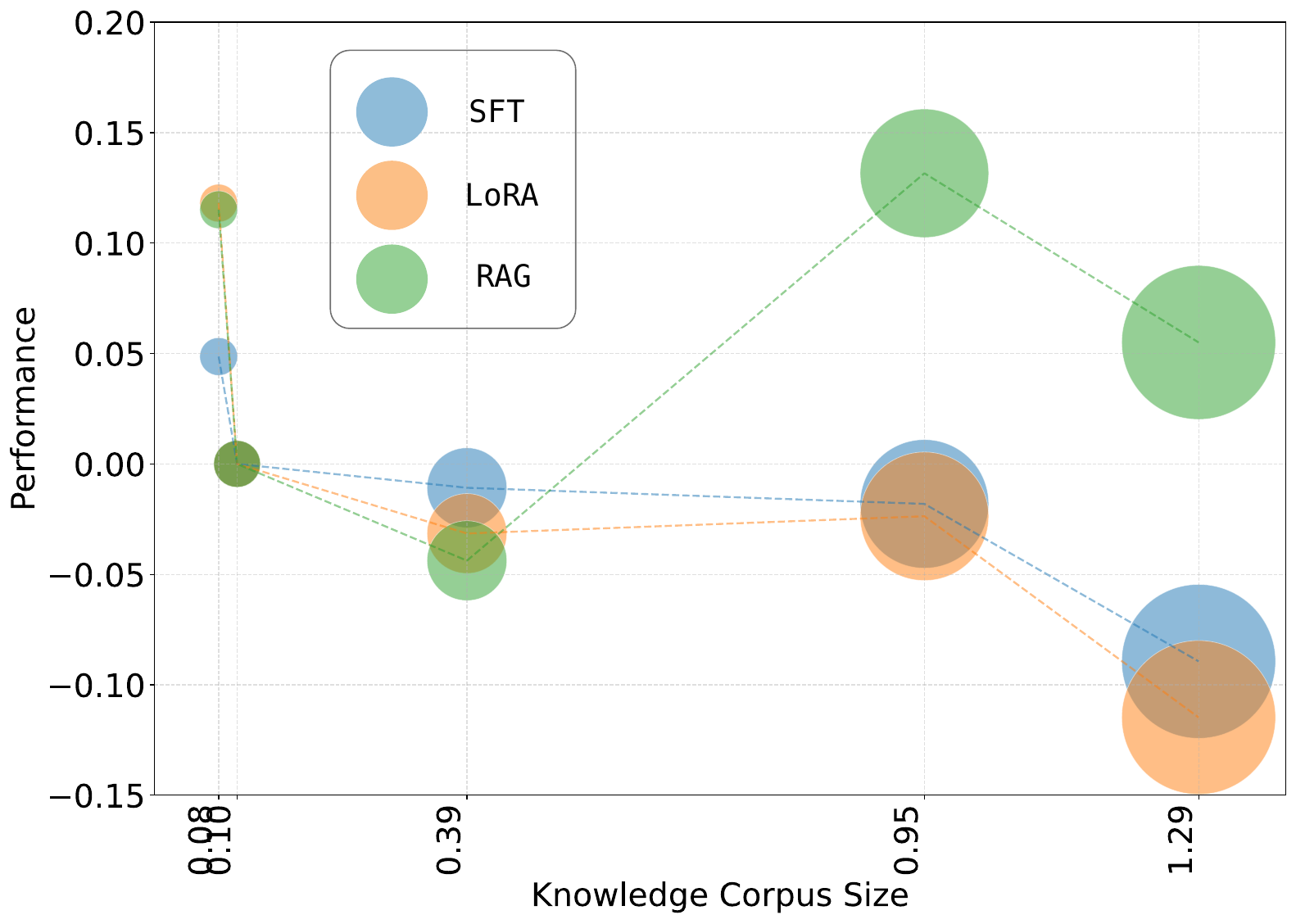} 
    \caption{The impact of knowledge corpus size on domain specialization performance.} 
    \label{fig:corpus size} 
\end{figure}

\paragraph{Impact of Knowledge Corpus Size.}
Understanding the relationship between knowledge corpus size and learning effectiveness can provide insights into data requirements for domain specialization.
We analyze this relationship on \ourapproach using SFT, LoRA, and RAG. Figure \ref{fig:corpus size} presents this analysis, where the x-axis represents knowledge corpus size measured in millions of tokens (the standard metric for LLM input), the y-axis shows AvgPassRate on the corresponding domain code generation tasks, and the point size indicates Lines of Code (LOC), a traditional software engineering metric for codebase scale. The results reveal distinct scaling behaviors across methods. Both SFT and LoRA exhibit a negative correlation between corpus size and performance, indicating that incorporating more domain-specific data paradoxically hinders learning effectiveness. In contrast, RAG demonstrates no such correlation, with its performance remaining largely unaffected by corpus size.

\begin{figure}[h!] 
    \centering 
    \includegraphics[width=0.6\textwidth]{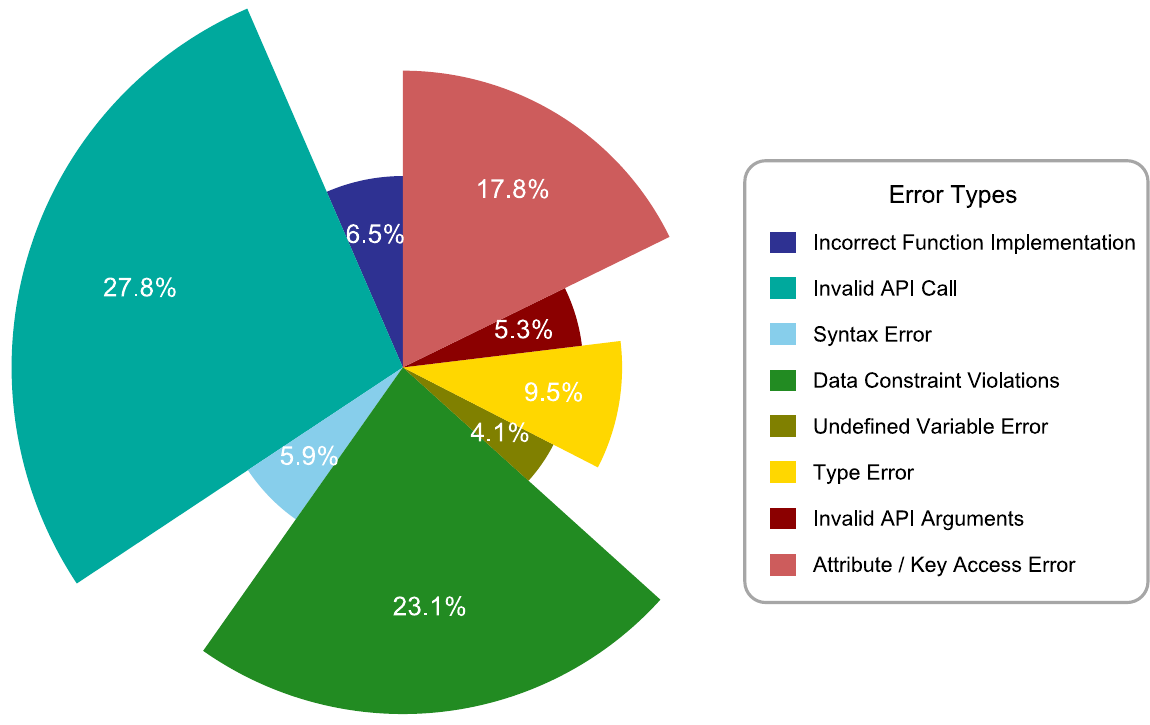} 
    \caption{Error type distribution of the base model on \ourapproach.} 
    \label{fig:error_analysis} 
\end{figure}

\paragraph{Error Analysis.}

To identify the key challenges posed by \ourapproach, we categorize and analyze the errors produced by base model. The distribution is presented in Figure~\ref{fig:error_analysis} (Appendix).
The largest error category is Invalid API Call, where LLMs hallucinate non-existent APIs. Combined with Invalid API Arguments, API-related errors account for approximately one-third of all failures. This indicates that current LLMs lack sufficient knowledge of framework-specific APIs, often falling back on generic patterns that do not align with the design of the framework. Data Constraint Violations represent the second largest category, where generated code fails to adhere to expected data formats, value ranges, or structural requirements. Together with Attribute/Key Access Error, Type Error, and Undefined Variable Error, these errors reveal that LLMs struggle to reason about the constraints governing data flow in domain-specific code. Other error types, such as Syntax Error and Incorrect Function Implementation represent a smaller share of the overall distribution.

\section{Conclusion}

We present \ourapproach, a benchmark that innovatively pairs domain knowledge corpora with evaluation tasks to assess how effectively LLMs can acquire and apply new domain knowledge, which enables systematic evaluation across domain code generation and knowledge understanding tasks, spanning 6 emerging domains with 11 frameworks.
Our experiments reveal that even SOTA LLMs struggle with domain code generation, and existing domain specialization methods offer only marginal improvements. Learning-based approaches suffer from diminishing returns as the knowledge corpus scales and exhibit catastrophic forgetting during continual learning, while agentic methods still remain far from meeting practical domain requirements. These findings highlight domain-specific software development as a significant challenge for LLMs, and we hope \ourapproach~will catalyze future research toward effective solutions.

\section*{Acknowledgments}
We thank Lecheng Wang and Liye Zhu for their assistance with the preliminary domain investigation. We are grateful to Zejun Wang for participation in discussions regarding test suite creation. We also acknowledge every member of the SEKE Team for their valuable suggestions provided during team discussions. This research is supported by the National Natural Science Foundation of China under Grant No. 62192733, 62192730, 62192731, the National Key R\&D Program under Grant No. 2023YFB4503801, and the Beijing Major Science and Technology Project under Contract No. Z251100008425005.

\bibliography{preprint}
\bibliographystyle{preprint}

\newpage
\appendix

\section{Structure and Statistics of \ourapproach}
\label{Structure and Statistics}

Table~\ref{tab:domain_frameworks} summarizes the hierarchical structure of \ourapproach. The benchmark is organized into two tasks: Domain Code Generation and Domain Knowledge Understanding. These tasks cover diverse domains including Reinforcement Learning, Agent, RAG, Model Optimization, Embodied AI, and Ascend Ecosystem. Each domain is grounded in one or more representative frameworks, and for the code generation task, we curate real-world projects built on these frameworks as evaluation targets.

\begin{table*}[h!]
\centering
\caption{Structure and statistics of \ourapproach across tasks, domains, and frameworks.}
\label{tab:domain_frameworks}
\resizebox{\textwidth}{!}{%
\begin{tabular}{p{1.7cm} p{1.5cm} p{2cm} p{3cm} p{2.5cm} p{1.8cm} p{3.5cm} p{4cm}}
\toprule
\textbf{Tasks} & \multicolumn{5}{c}{\textbf{Domain Code Generation}} 
 & \multicolumn{2}{c}{\textbf{Domain Knowledge Understanding}} \\
\cmidrule(lr){2-6} \cmidrule(lr){7-8}
\textbf{Domain} 
& \multicolumn{2}{c}{Reinforcement Learning} 
& Agent 
& RAG 
& Model Optimization 
& Embodied AI 
& Ascend Ecosystem \\
\cmidrule(lr){2-3}\cmidrule(lr){4-4}\cmidrule(lr){5-5}\cmidrule(lr){6-6}\cmidrule(lr){7-7}\cmidrule(lr){8-8}
\textbf{Framework} 
& VeRL 
& Open-R1 
& Smolagents 
& RAG-anything 
& TensorRT 
& VSLAM-LAB, cosmos-rl, robocasa, trackerLab 
& ascend-transformer-boost, triton-ascend \\
\cmidrule(lr){2-2}\cmidrule(lr){3-3}\cmidrule(lr){4-4}\cmidrule(lr){5-5}\cmidrule(lr){6-6}\cmidrule(lr){7-7}\cmidrule(lr){8-8}
\textbf{Project} 
& prime, PURE, ARES, LUFFY, DAPO, PACS, critic-rl
& AlphaDrive, CANOE, VLM-R1, VisualQuality-R1, ml-diffucoder
& DeepSearchAgents, ToolBrain, examples, smolcc
& BookWorm, CHAT-ANYTHING, law-unleashed-rag, obsidianGraphRAG, rag4chat
& FlagScale, Nemo, TensorRT-Incubator, byteperf
& \makecell{--} 
& \makecell{--}  \\
\bottomrule
\end{tabular}%
}
\end{table*}

\section{Performance Analysis of Project Code Generation}\label{sec:project_level_generation}

Given the high task complexity and prohibitive computational cost, we perform a qualitative analysis on the project-level code generation task of \ourapproach. Given the absence of established methods for truly project code generation, we adopt Claude Code (based on Claude Sonnet 4.5), a state-of-the-art coding agent, augmented with the VeRL framework as an explicit domain knowledge base, to tackle this task.
Given the project description and module division specification, the agent is tasked with generating a complete project that passes integration tests.

After generation, we execute the produced project end-to-end using a faithful debugging protocol: runtime errors are logged, minimally fixed, and the project is re-executed until the training pipeline runs successfully. This procedure allows us to analyze the types of errors that arise during execution.

We observe that most failures stem from a misalignment between the generated code and implicit domain knowledge. While the agent successfully reproduces the overall project structure and reuses framework-style components, it frequently violates unstated assumptions that are only enforced at runtime. These failures can be categorized into two main types:

\begin{itemize}
\item \textbf{Interface Misalignment.} The agent generates components that are individually correct but incompatible when composed together. This includes mismatches between data formats, configuration semantics, and module interfaces, suggesting that the agent lacks understanding of cross-module contracts within the framework.

\item \textbf{Execution Semantics Misunderstanding.} The agent produces code that is syntactically valid but violates the runtime execution logic of the framework, particularly in training dynamics. This indicates that the agent fails to capture how different components interact during actual execution.
\end{itemize}

These findings suggest that the agent relies primarily on surface-level pattern imitation rather than a deep understanding of the execution logic of the framework. Effective project-level code generation requires not only structural knowledge of the framework, but also implicit knowledge about runtime behaviors and cross-module dependencies. Advancing project-level code generation remains an open challenge, and we hope \ourapproach provides a meaningful benchmark for future research in this direction.

\section{Performance of Code-specific Domain Specialization Methods}
Since KNM-LM~\citep{tang2023knm-lm} and DSCC~\citep{yu2025enhancing} are originally designed for code completion, we carefully adapt them to code generation for evaluation. Results are shown in Appendix Table~\ref{tab:code_domain_specialization_methods}.

\begin{table*}[h]
\centering
\caption{Performance of domain specialization methods for code on \ourapproach.}
\label{tab:code_domain_specialization_methods}
\resizebox{\textwidth}{!}{%
\begin{tabular}{lcccccccccc}
\toprule
\cmidrule(lr){2-11}
\multirow{2}{*}{Method}& \multicolumn{2}{c}{RL} 
& \multicolumn{2}{c}{Agent} 
& \multicolumn{2}{c}{RAG} 
& \multicolumn{2}{c}{\makecell{Model\\Optimization}} 
& \multicolumn{2}{c}{\cellcolor{gray!20}Average} \\
\cmidrule(lr){2-3} \cmidrule(lr){4-5} \cmidrule(lr){6-7} \cmidrule(lr){8-9} \cmidrule(lr){10-11}
& Pass@1 & APR 
& Pass@1 & APR 
& Pass@1 & APR 
& Pass@1 & APR 
& \cellcolor{gray!20}Pass@1 & \cellcolor{gray!20}APR \\
\midrule
Base Model 
& 3.8 & 11.4 
& 3.9 & 17.1 
& 0.0 & 0.0 
& 22.2 & 33.4 
& \cellcolor{gray!20}7.5 & \cellcolor{gray!20}15.5 \\
kNM-LM 
& 1.8 & 3.3 
& 0.0 & 0.0 
& 0.0 & 0.0 
& 5.6 & 10.5 
& \cellcolor{gray!20}1.8 & \cellcolor{gray!20}3.4 \\
DSCC 
& 3.6 & 11.1 
& 1.9 & 17.5 
& 0.0 & 0.0 
& 22.2 & 37.1 
& \cellcolor{gray!20}6.9 & \cellcolor{gray!20}13.1 \\
\bottomrule
\end{tabular}%
}
\end{table*}

\section{Additional Domain Specialization Baselines}\label{sec:additional_baselines}

To further investigate the effectiveness of domain specialization methods beyond those reported in the main paper, we evaluate additional baselines including dense retrieval-based RAG and instruction-tuned fine-tuning with synthesized training pairs.

\textbf{Dense RAG.} We replace the BM25 lexical retriever with an embedding-based dense retriever (EmbeddingGemma), keeping all other RAG configurations identical to the original RAG baseline.

\textbf{Instruction-tuned Fine-tuning.} Instead of training on raw corpus text via next-token prediction, we use Gemini-3-Flash to generate code generation requirements for functions from our knowledge corpus, creating training pairs of (requirement, code) for instruction-tuned fine-tuning.

The results are presented in Table~\ref{tab:additional_baselines}. For reference, we also include the original SFT and RAG baselines from the main paper. These additional approaches yield marginal improvements, with average performance across domains remaining below 10\% Pass@1, which does not affect our original findings that existing domain specialization methods provide only limited gains.

\begin{table*}[h]
\centering
\caption{Performance of additional domain specialization baselines on \ourapproach.}
\label{tab:additional_baselines}
\resizebox{\textwidth}{!}{%
\begin{tabular}{lcccccccccc}
\toprule
\multirow{2}{*}{Method}& \multicolumn{2}{c}{RL}
& \multicolumn{2}{c}{Agent}
& \multicolumn{2}{c}{RAG}
& \multicolumn{2}{c}{\makecell{Model\\Optimization}}
& \multicolumn{2}{c}{\cellcolor{gray!20}Average} \\
\cmidrule(lr){2-3} \cmidrule(lr){4-5} \cmidrule(lr){6-7} \cmidrule(lr){8-9} \cmidrule(lr){10-11}
& Pass@1 & APR
& Pass@1 & APR
& Pass@1 & APR
& Pass@1 & APR
& \cellcolor{gray!20}Pass@1 & \cellcolor{gray!20}APR \\
\midrule
SFT
& 7.3 & 12.9
& 1.9 & 16.1
& 0.0 & 0.0
& 16.7 & 24.4
& \cellcolor{gray!20}6.5 & \cellcolor{gray!20}13.4 \\
RAG (BM25)
& 7.2 & 23.7
& 5.8 & 12.8
& 0.0 & 0.0
& 16.7 & 38.9
& \cellcolor{gray!20}7.4 & \cellcolor{gray!20}18.9 \\
\cdashline{1-11}
Dense RAG
& 7.3 & 18.4
& 5.8 & 12.4
& 0.0 & 0.0
& 22.2 & 45.0
& \cellcolor{gray!20}8.8 & \cellcolor{gray!20}18.9 \\
Instruction-tuned
& 5.4 & 15.4
& 13.5 & 35.2
& 0.0 & 0.0
& 16.7 & 27.0
& \cellcolor{gray!20}8.9 & \cellcolor{gray!20}19.4 \\
\bottomrule
\end{tabular}%
}
\end{table*}

\section{Ablation on Maximum Sequence Length}\label{sec:max_seq_len_ablation}

The default SFT configuration uses a maximum sequence length of 2048 tokens, while the knowledge corpora average approximately 77K lines. To verify whether the maximum sequence length is a determining factor, we conduct an ablation study on the SFT baseline with three different maximum sequence lengths: 2048, 4096, and 8192. As shown in Table~\ref{tab:max_seq_len}, the results demonstrate that maximum sequence length has minimal impact on SFT performance and is not a determining factor for knowledge acquisition.

\begin{table*}[h]
\centering
\caption{Ablation study on maximum sequence length for SFT on \ourapproach.}
\label{tab:max_seq_len}
\resizebox{\textwidth}{!}{%
\begin{tabular}{lcccccccccc}
\toprule
\multirow{2}{*}{Method}& \multicolumn{2}{c}{RL}
& \multicolumn{2}{c}{Agent}
& \multicolumn{2}{c}{RAG}
& \multicolumn{2}{c}{\makecell{Model\\Optimization}}
& \multicolumn{2}{c}{\cellcolor{gray!20}Average} \\
\cmidrule(lr){2-3} \cmidrule(lr){4-5} \cmidrule(lr){6-7} \cmidrule(lr){8-9} \cmidrule(lr){10-11}
& Pass@1 & APR
& Pass@1 & APR
& Pass@1 & APR
& Pass@1 & APR
& \cellcolor{gray!20}Pass@1 & \cellcolor{gray!20}APR \\
\midrule
SFT-2048
& 7.3 & 12.9
& 1.9 & 16.1
& 0.0 & 0.0
& 16.7 & 24.4
& \cellcolor{gray!20}6.5 & \cellcolor{gray!20}13.4 \\
SFT-4096
& 5.6 & 15.7
& 1.9 & 19.9
& 0.0 & 0.0
& 16.7 & 29.8
& \cellcolor{gray!20}6.0 & \cellcolor{gray!20}16.3 \\
SFT-8192
& 3.8 & 15.4
& 1.9 & 19.9
& 0.0 & 0.0
& 22.2 & 30.6
& \cellcolor{gray!20}7.0 & \cellcolor{gray!20}16.5 \\
\bottomrule
\end{tabular}%
}
\end{table*}

\section{Pass@any Results}\label{sec:pass_any_results}

To explore the performance upper bound, we report Pass@any results, which considers a sample correct if at least one of 10 attempts (with temperature=0.8) produces a correct solution. Table~\ref{tab:pass_any_llm} presents Pass@any results for various LLMs, and Table~\ref{tab:pass_any_methods} presents results for domain specialization methods. Pass@any shows moderate improvements over Pass@1, providing insights into the performance ceiling.

\begin{table*}[h]
\centering
\caption{Pass@any results of LLMs on \ourapproach (domain code generation).}
\label{tab:pass_any_llm}
\resizebox{\textwidth}{!}{%
\begin{tabular}{lcccccccccc}
\toprule
\multirow{2}{*}{Models}& \multicolumn{2}{c}{RL}
& \multicolumn{2}{c}{Agent}
& \multicolumn{2}{c}{RAG}
& \multicolumn{2}{c}{\makecell{Model\\Optimization}}
& \multicolumn{2}{c}{\cellcolor{gray!20}Average} \\
\cmidrule(lr){2-3} \cmidrule(lr){4-5} \cmidrule(lr){6-7} \cmidrule(lr){8-9} \cmidrule(lr){10-11}
& Pass@1 & Pass@any
& Pass@1 & Pass@any
& Pass@1 & Pass@any
& Pass@1 & Pass@any
& \cellcolor{gray!20}Pass@1 & \cellcolor{gray!20}Pass@any \\
\midrule
Claude-Sonnet-4-5
& 8.2 & 9.8
& 12.7 & 17.3
& 0.0 & 0.0
& 16.7 & 16.7
& \cellcolor{gray!20}9.4 & \cellcolor{gray!20}10.9 \\
Kimi-K2-Instruct
& 8.4 & 14.9
& 14.8 & 19.2
& 0.0 & 0.0
& 19.4 & 22.2
& \cellcolor{gray!20}10.7 & \cellcolor{gray!20}14.1 \\
DeepSeek-V3.1
& 5.8 & 11.4
& 14.2 & 19.2
& 0.0 & 0.0
& 16.7 & 16.7
& \cellcolor{gray!20}9.2 & \cellcolor{gray!20}11.8 \\
GPT-5-mini
& 10.2 & 13.8
& 13.1 & 19.2
& 0.0 & 0.0
& 16.7 & 16.7
& \cellcolor{gray!20}10.0 & \cellcolor{gray!20}12.4 \\
\cdashline{1-11}
Qwen2.5-Coder-32B-Instruct
& 1.8 & 1.8
& 3.7 & 9.6
& 0.0 & 0.0
& 5.6 & 5.6
& \cellcolor{gray!20}2.7 & \cellcolor{gray!20}4.2 \\
Qwen2.5-Coder-7B-Instruct
& 2.9 & 7.6
& 3.7 & 3.9
& 0.0 & 0.0
& 14.4 & 22.2
& \cellcolor{gray!20}5.3 & \cellcolor{gray!20}8.4 \\
Llama-3.1-8B-Instruct
& 3.6 & 11.6
& 5.4 & 15.4
& 0.0 & 0.0
& 16.7 & 16.7
& \cellcolor{gray!20}6.4 & \cellcolor{gray!20}10.9 \\
\bottomrule
\end{tabular}%
}
\begin{flushleft}
\tiny
Pass@1 values here are computed from temperature=0.8 sampling (10 samples), which may differ slightly from the greedy-decoded Pass@1 in Table~\ref{tab:LLMs}.
\end{flushleft}
\end{table*}

\begin{table*}[h]
\centering
\caption{Pass@any results of domain specialization methods on \ourapproach (domain code generation).}
\label{tab:pass_any_methods}
\resizebox{\textwidth}{!}{%
\begin{tabular}{lcccccccccc}
\toprule
\multirow{2}{*}{Method}& \multicolumn{2}{c}{RL}
& \multicolumn{2}{c}{Agent}
& \multicolumn{2}{c}{RAG}
& \multicolumn{2}{c}{\makecell{Model\\Optimization}}
& \multicolumn{2}{c}{\cellcolor{gray!20}Average} \\
\cmidrule(lr){2-3} \cmidrule(lr){4-5} \cmidrule(lr){6-7} \cmidrule(lr){8-9} \cmidrule(lr){10-11}
& Pass@1 & Pass@any
& Pass@1 & Pass@any
& Pass@1 & Pass@any
& Pass@1 & Pass@any
& \cellcolor{gray!20}Pass@1 & \cellcolor{gray!20}Pass@any \\
\midrule
Base Model
& 2.9 & 7.6
& 3.7 & 3.9
& 0.0 & 0.0
& 14.4 & 22.2
& \cellcolor{gray!20}5.3 & \cellcolor{gray!20}8.4 \\
SFT
& 3.4 & 7.3
& 7.3 & 17.3
& 0.0 & 0.0
& 15.6 & 22.2
& \cellcolor{gray!20}6.6 & \cellcolor{gray!20}11.7 \\
LoRA
& 1.8 & 7.2
& 5.8 & 11.5
& 0.0 & 0.0
& 14.4 & 22.2
& \cellcolor{gray!20}5.5 & \cellcolor{gray!20}10.2 \\
RAG
& 4.0 & 9.1
& 6.5 & 21.2
& 0.0 & 0.0
& 17.2 & 27.8
& \cellcolor{gray!20}6.9 & \cellcolor{gray!20}14.5 \\
kNN-LM
& 5.2 & 5.4
& 6.5 & 9.6
& 0.0 & 0.0
& 16.7 & 16.7
& \cellcolor{gray!20}7.1 & \cellcolor{gray!20}7.9 \\
\bottomrule
\end{tabular}%
}
\begin{flushleft}
\tiny
Pass@1 values here are computed from temperature=0.8 sampling (10 samples), which may differ slightly from the greedy-decoded Pass@1 in Table~\ref{tab:domain_specialization_methods}.
\end{flushleft}
\end{table*}

\section{Control Experiments for Agent Evaluation}\label{sec:control_agent}

In Table~\ref{tab:detail_agent}, we controlled for the base model across open-source agents (using Qwen2.5-Coder-32B-Instruct for both SWE-Agent and OpenHands, whose training data cutoff predates our dataset's cutoff date to ensure contamination-free evaluation), while Claude Code is powered by Claude-Sonnet-4-5. Since Claude Code's terms of service restrict its use to Claude models only, this introduces a confounding factor. To address this confound, we conduct control experiments: Claude-Sonnet-4-5 + SWE-Agent and Claude-Sonnet-4-5 + OpenHands. The results are presented in Table~\ref{tab:control_agent}.

The results show that switching to the stronger Claude-Sonnet-4-5 base model yields notable improvements with both SWE-Agent and OpenHands, yet their performance still falls substantially short of Claude Code. These results reveal a gap between current open-source agents and closed-source agents, indicating that further optimization is needed for open-source agents, which does not affect our original conclusion.

\begin{table*}[h]
\centering
\caption{Control experiments: performance of agents with Claude-Sonnet-4-5 as the base model on \ourapproach.}
\label{tab:control_agent}
\resizebox{\textwidth}{!}{%
\begin{tabular}{lcccccccccc}
\toprule
\multirow{2}{*}{Method}& \multicolumn{2}{c}{RL}
& \multicolumn{2}{c}{Agent}
& \multicolumn{2}{c}{RAG}
& \multicolumn{2}{c}{\makecell{Model\\Optimization}}
& \multicolumn{2}{c}{\cellcolor{gray!20}Average} \\
\cmidrule(lr){2-3} \cmidrule(lr){4-5} \cmidrule(lr){6-7} \cmidrule(lr){8-9} \cmidrule(lr){10-11}
& Pass@1 & APR
& Pass@1 & APR
& Pass@1 & APR
& Pass@1 & APR
& \cellcolor{gray!20}Pass@1 & \cellcolor{gray!20}APR \\
\midrule
Claude-Sonnet-4-5 (Base)
& 7.6 & 28.6
& 5.8 & 21.3
& 0.0 & 0.0
& 11.1 & 34.8
& \cellcolor{gray!20}6.1 & \cellcolor{gray!20}21.2 \\
\cdashline{1-11}
SWE-Agent
& 9.1 & 26.2
& 28.9 & 46.2
& 25.0 & 25.0
& 11.1 & 25.1
& \cellcolor{gray!20}18.5 & \cellcolor{gray!20}30.6 \\
OpenHands
& 10.9 & 28.3
& 17.3 & 30.5
& 25.0 & 25.0
& 16.7 & 34.2
& \cellcolor{gray!20}17.5 & \cellcolor{gray!20}29.5 \\
Claude Code
& 13.6 & 26.4
& 38.5 & 56.3
& 62.5 & 68.8
& 22.2 & 45.6
& \cellcolor{gray!20}34.2 & \cellcolor{gray!20}49.3 \\
\bottomrule
\end{tabular}%
}
\end{table*}

\section{Data Contamination Detection in \ourapproach} \label{sec:data_contamination_detection}
To verify whether our dataset suffers from data contamination, we apply CDD~\citep{dong2024cdd}, a widely adopted contamination detection method, to the base models evaluated in our experiments. The contamination index quantifies the likelihood that an LLM has been exposed to benchmark data during training, ranging from 0 to 1, where lower values indicate less contamination.
We conduct detection on both Knowledge Corpus and Test Set across all domains, with results shown in Table~\ref{tab:data_contamination}. The average contamination index for Knowledge Corpus is 0.08, while that for Test Set is as low as 0.005, providing strong evidence that our benchmark is free from data contamination.

\begin{table*}[htbp]
\centering
\caption{Data contamination index on \ourapproach, where DCI indicates Data Contamination Index.}
\label{tab:data_contamination}
\resizebox{\textwidth}{!}{
\begin{tabular}{lcccccccccc}
\toprule
\multirow{2}{*}{\textbf{Metric}}
 & \multicolumn{2}{c}{RL} 
 & \multicolumn{2}{c}{Agent} 
 & \multicolumn{2}{c}{RAG} 
 & \multicolumn{2}{c}{Model Optimization} 
 & \multicolumn{2}{c}{Average} \\
\cmidrule(lr){2-3}
\cmidrule(lr){4-5}
\cmidrule(lr){6-7}
\cmidrule(lr){8-9}
\cmidrule(lr){10-11}
 & Knowledge Corpus & Test Set 
 & Knowledge Corpus & Test Set 
 & Knowledge Corpus & Test Set 
 & Knowledge Corpus & Test Set 
 & Knowledge Corpus & Test Set \\
\midrule
\makecell[l]{DCI}
& 0.037 & 0.000 
& 0.050 & 0.020 
& 0.154 & 0.000  
& 0.080 & 0.000 
& 0.080 & 0.005 \\
\bottomrule
\end{tabular}
}
\end{table*}

\section{Performance-Cost Trade-off Analysis}\label{sec:performance_cost_trade_offs}

\begin{figure}[h] 
    \centering 
    \includegraphics[width=0.6\textwidth]{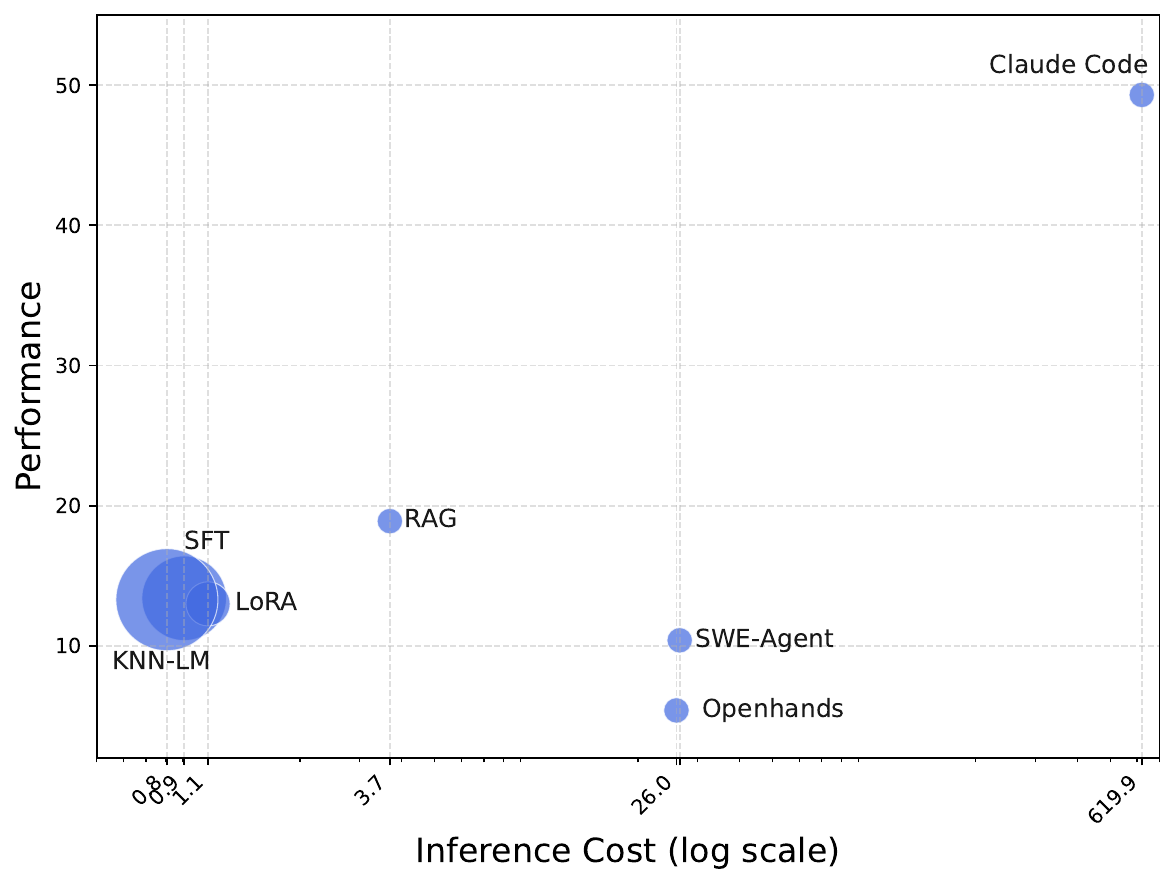} 
    \caption{Results of Performance-Cost Trade-off.} 
    \label{fig:performance_cost_trade_offs} 
\end{figure}

To provide a comprehensive understanding of the trade-offs between effectiveness and efficiency, we analyze both the training and inference costs of knowledge learning and utilization methods, as well as coding agents.
Figure \ref{fig:performance_cost_trade_offs}  illustrates the performance-cost landscape of different approaches. The y-axis represents performance (measured by Pass@1), while the x-axis shows inference token consumption. The size of each scatter point indicates the training time required by the method. This analysis provides actionable guidance for practitioners to identify suitable methods given their specific resource constraints.

\section{Case Study}\label{sec:case_study}
\label{appendix:cases}

We provide two cases to illustrate the most frequent error types identified in our error analysis: Invalid API Call and Data Constraint Violations.

\begin{figure*}[t!] 
    \centering 
    \includegraphics[width=0.98\textwidth]{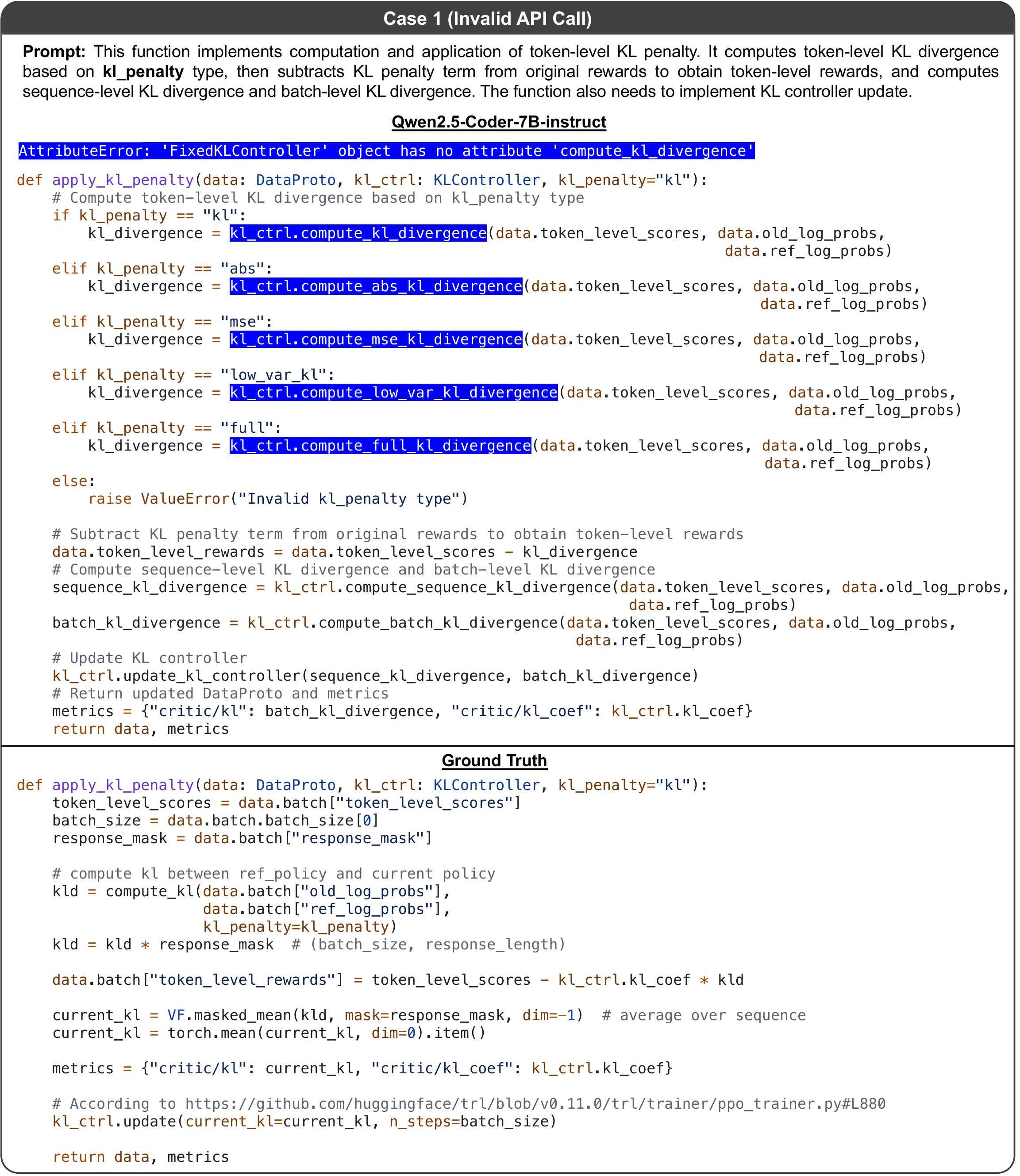} 
    \caption{A case of invalid API call.} 
    \label{fig:case1} 
\end{figure*}

\textbf{Case 1: Invalid API Call.} Figure~\ref{fig:case1} presents an example of an Invalid API Call error. In this case, the model hallucinates a series of non-existent APIs, including \texttt{compute\_kl\_divergence}, \texttt{compute\_abs\_kl\_divergence}, \texttt{compute\_mse\_kl\_divergence}, \texttt{compute\_low\_var\_kl\_divergence}, and \texttt{compute\_full\_kl\_divergence}. This error arises because the model is unfamiliar with the \texttt{FixedKLController} class defined in the knowledge corpus, and instead fabricates plausible-looking but invalid method names based on surface-level patterns.

\begin{figure*}[t!] 
    \centering 
    \includegraphics[width=0.98\textwidth]{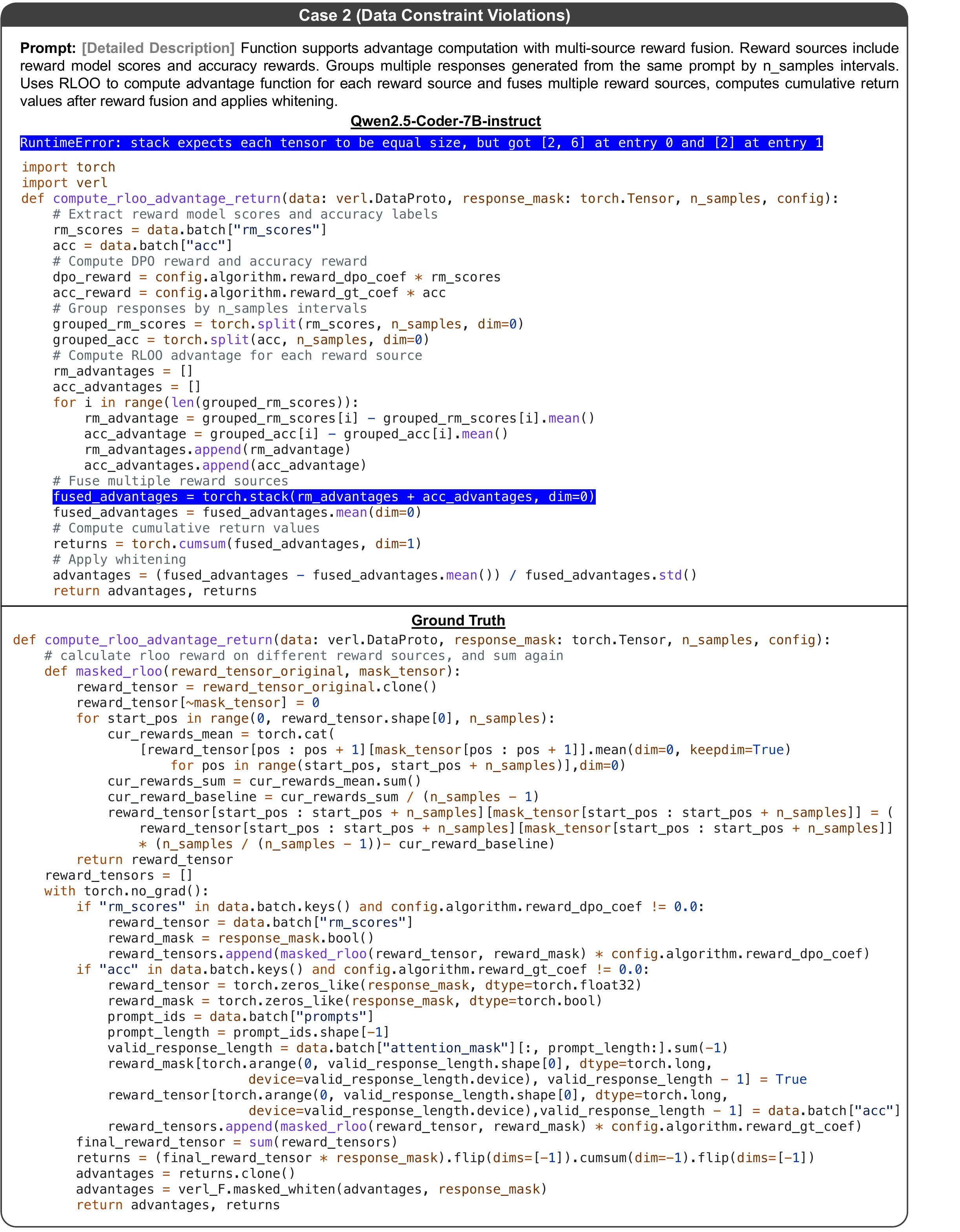} 
    \caption{A case of data constraint violations.} 
    \label{fig:case2} 
\end{figure*}

\textbf{Case 2: Data Constraint Violations.} Figure~\ref{fig:case2} illustrates an example of a Data Constraint Violations error. In this case, the model is unaware of the implicit convention of the framework that \texttt{rm\_scores} operates at token-level granularity while \texttt{acc} is defined at sequence-level. Due to this lack of understanding, the model incorrectly fuses rewards of different granularities as homogeneous objects and projects them onto an inconsistent temporal scale for RLOO computation, resulting in semantic errors during execution.

\section{Additional Implementation Details} \label{sec:additional_implementation_details}

\subsection{Experiment Setup} \label{sec:experiment_setup}

\textbf{Domain Code Generation}. We primarily evaluate the generation of core functions. Since generating entire projects proves extremely challenging (which we present in Appendix B), we focus on core function generation as the main evaluation target.
For direct LLM generation, we provide the core function description and function signature as input, prompting models to generate the function body. For knowledge learning and utilization methods, we distinguish between two scenarios: 1) Inference-based methods can access both the knowledge corpus and the instance project context, but we carefully remove the target core functions to prevent data leakage. 2) Training-based methods are trained on the knowledge corpus.

For the domain code generation task, we employ Pass@1 as the primary metric, measuring the proportion of samples where all test cases pass. Since Pass@1 is quite stringent, we also adopt a more lenient metric, AvgPassRate (APR), which measures the average proportion of test cases passed per sample. Additionally, to explore the performance upper bound of models and methods, we use Pass@any, which considers a sample correct if at least one of multiple attempts produces a correct solution. In practice, we generate 10 samples for Pass@any.

\textbf{Domain Knowledge Understanding}. Models are provided with questions and answer options and are asked to select the correct answer(s). Knowledge learning and utilization methods can leverage the knowledge corpus when answering. We use Accuracy (ACC) as the evaluation metric for this task.

By default, we employ greedy decoding (temperature=0) for all evaluations. The only exception is Pass@any, which requires sampling diversity and therefore uses temperature=0.8. 

In Table \ref{tab:LLMs}, we evaluate 10 well-known LLMs on both Domain Code Generation and Domain Knowledge Understanding tasks, including both closed-source and open-source models across different model families and parameter scales. For closed-source models, we adopt Gemini-2.5-pro~\citep{Gemini3}, Claude-Sonnet-4.5~\citep{claudecode45}, GPT-5-mini~\citep{gpt-5-mini} and o4-mini~\citep{o4-mini}. For open-source models, we adopt Kimi-K2-Instruct~\citep{kimi}, DeepSeek-V3.1~\citep{deepseek-v3.1}, Qwen2.5-Coder-32B-Instruct~\citep{qwen-32b}, Qwen2.5-Coder-7B-Instruct~\citep{qwen-7b}, Llama-3.1-8B-Instruct~\citep{llama-8b} and Deepseek-Coder-7b-Base~\citep{deepseek-7b}.

\subsection{Setup of Domain Specialization Methods}
We provide the implementation details of domain specialization methods evaluated in our experiments.

\textbf{SFT}~\citep{dong2024abilities} We perform full-parameter fine-tuning based on the Next-Token Prediction (NTP) objective. We use a learning rate of $5 \times 10^{-6}$ with cosine decay, train for 5 epochs with a maximum sequence length of 2048, and use BF16 precision.

\textbf{LoRA}~\citep{lora} We adopt the PEFT framework, freezing pre-trained weights and capturing domain-specific knowledge through low-rank decomposition matrices. We set the learning rate of $1 \times 10^{-4}$, rank $r=16$, alpha $\alpha=32$, and dropout to 0.05, training for 5 epochs.

\textbf{RAG}~\citep{lewis2020retrieval} We employ BM25 for lexical retrieval. The knowledge corpus is chunked at both function-level and class-level granularities, covering instance-specific context and framework-wide knowledge. We retrieve top-5 results using queries constructed from function names and parameter signatures.

\textbf{kNN-LM}~\citep{knn-lm}, \textbf{kNM-LM}~\citep{tang2023knm-lm}, and \textbf{DSCC}~\citep{yu2025enhancing}. We use the official implementations and default configurations provided in the original papers. We adapt these methods to our tasks by modifying the input-output format without altering their core algorithms.

\textbf{Training Data Organization for SFT and LoRA.} Both SFT and LoRA follow the same data organization procedure, consistent with standard practices used in modern code LLMs such as StarCoder and DeepSeek-Coder~\citep{deepseek-7b}. Specifically, we concatenate all files from the knowledge corpus (raw documentation, source code, examples, etc.) in random order, with special separators \texttt{<|endoftext|>} inserted between files. The concatenated text is then split into equal-length training chunks based on the maximum context length. All chunks participate in training, ensuring complete coverage of the knowledge corpus.

\subsection{Setup of LLM Agents}
We evaluate three agentic coding systems on \ourapproach: SWE-Agent~\citep{yang2024sweagent}, OpenHands~\citep{wang2025openhands}, and Claude Code~\citep{claudecode}. SWE-Agent is configured with five exploration tools (\texttt{find\_file}, \texttt{search\_dir}, \texttt{search\_file}, \texttt{view\_code}, \texttt{filemap}). We set max\_turns=30 and timeout=3000s per task. OpenHands uses the CodeAct agent with Terminal Tool for bash execution and file operations. We limit execution to max\_steps=30 per task. Claude Code is invoked via CLI, allowing \texttt{Read/Write/Edit/Grep/Glob/Bash} tools. We set max\_turns=30 and timeout=3000s per task, requiring structured JSON output for reliable result extraction.

\subsection{Experiment Setup of Investigating Knowledge Forgetting and Conflict} 
Understanding these is essential for developing effective continual learning strategies in domain-specific software development.
We investigate this problem on \ourapproach using SFT, which offers a controlled experimental setting to isolate the effects of knowledge accumulation without confounding factors from retrieval or auxiliary inference mechanisms. We compare three knowledge learning scenarios: (1) \textit{Single Framework} (default): train on the knowledge corpus of one framework and evaluate on projects built with that framework, (2) \textit{Multi-Framework, Same Domain}: sequentially train on two frameworks within the same domain (e.g., two RL frameworks), then evaluate on the first-learned framework to measure intra-domain knowledge retention, and (3) \textit{Multi-Framework, Cross-Domain}: sequentially train on two frameworks from distinct domains (e.g., a RL framework followed by a agent framework), then evaluate on the first-learned framework to assess cross-domain interference.

\section{Limitations}

Our work has the following two limitations: 
First, since LLMs continue to evolve, we take proactive measures to mitigate data contamination risks and ensure reliable evaluation. We explicitly establish a dataset cutoff date and select frameworks created after this date. When evaluating domain specialization methods, we deliberately use base models whose training cutoff precedes the dataset's cutoff. Additionally, we conduct contamination detection (detailed in Appendix \ref{sec:data_contamination_detection}), which demonstrates negligible contamination probability. The challenging nature of \ourapproach, where even SOTA models achieve limited performance, further validates the robustness of our benchmark against potential data contamination concerns for domain specialization methods.
Second, constrained by API availability and access limitations at the time of our experiments, we do not evaluate several recently released LLMs, such as OpenAI GPT-5 and Gemini 3 Pro. We will extend our evaluation to cover the emerging models in future updates of \ourapproach.

\end{document}